# Mechanobiology of shear-induced platelet aggregation leading to occlusive arterial thrombosis: a multiscale in silico analysis

Zixiang L Liu*[€], David N Ku[§], and Cyrus K Aidun[‡]

George W. Woodruff School of Mechanical Engineering, and Parker H. Petit Institute for Bioengineering and Biosciences, Georgia Institute of Technology, Georgia, United States 30332

**Abstract**

Occlusive thrombosis in arteries causes heart attacks and strokes. The rapid growth of thrombus at elevated shear rates (~10,000 1/s) relies on shear-induced platelet aggregation (SIPA) thought to come about from the entanglement of von Willebrand factor (VWF) molecules. The mechanism for SIPA is not yet understood in terms of cell- and molecule-level dynamics in fast flowing bloodstreams. Towards this end, we develop a multiscale computational model to recreate SIPA *in silico*, where the suspension dynamics and interactions of individual platelets and VWF multimers are resolved directly. The platelet-VWF interaction via GP1b-A1 bonds is prescribed with intrinsic binding rates theoretically derived and informed by single-molecule measurements. The model is validated against existing microfluidic SIPA experiments, showing good agreement with the *in vitro* observations in terms of the morphology, traveling distance and capture time of the platelet aggregates. Particularly, the capture of aggregates can occur in a few milliseconds, comparable to the platelet transit time through pathologic arterial stenotic sections and much shorter than the time for shear-induced platelet activation. The multiscale SIPA simulator provides a cross-scale tool for exploring the biophysical mechanisms of SIPA in silico that are difficult to access with single-molecule measurements or micro-/macro-fluidic assays only.

*Present address: Division of Applied Math, Brown University, Providence, RI 02912, USA
[€]Email: zixiang_liu@brown.edu
[§]Email: david.ku@me.gatech.edu
[‡]Email: cyrus.aidun@me.gatech.edu

## 1. Introduction

Occlusive arterial thrombosis is an acute pathological condition where a platelet-rich clot can rapidly form and occlude a stenotic artery, which is the direct cause of ischemic stroke and myocardial infarction as two major causes of death globally (Virani et al 2020). The arterial thrombus requires to grow against ultra-high shear stresses created by the fast-flowing blood through an atherosclerotic stenosis (Casa & Ku 2017, Ku et al 1985); see Fig. 1(a). The Virchow's triad for stagnant blood clotting fails to explain high-shear arterial thrombosis (Virchow 1862). Rather, it can be explained by an alternative triad, which includes a pro-thrombotic surface, pathologically high shear, and blood constituents mainly including platelets and von Willebrand factor (VWF) (Casa et al 2015).

It is now established that VWF plays a crucial role in supporting the growth of high-shear thrombus from different scales (Casa & Ku 2017, Kim et al 2019). At molecular scale, platelet mural adhesion under high shear ($>3000\ s^{-1}$) exclusively depends on VWF through rapid formation of GP1b-A1 bonds (Savage et al 1998). The GP1b-A1 kinetic rates have recently been measured using single-molecule measurements (Chen et al 2019, Fu et al 2017, Kim et al 2010, Zhang et al 2015). At macromolecular scale, the VWF multimer changes its tertiary structure from globular to elongated states and exposes more binding sites, when subjected to certain shear-rate thresholds depending on whether the VWF is suspended (~6000 $s^{-1}$) (Alexander-Katz et al 2006, Schneider et al 2007) or adhered (~1000 $s^{-1}$) (Fu et al 2017). At the cellular scale, VWF causes shear-induced platelet aggregation (SIPA) through GP1b-A1 bonds independent of platelet activation under elevated high shear ($>10,000\ s^{-1}$) (Jackson 2007, Ruggeri et al 2006).

Despite the identification of separate roles of VWF in arterial thrombosis at different scales, it remains elusive how such molecular and cellular ingredients can build up all the way to a micro-thrombus via SIPA. Such information is difficult to obtain experimentally as that requires imaging microsecond-scale events occurring in fast-flowing bloodstream with both molecular and cellular resolutions.

Computational models for arterial thrombosis have been developed with the hope of recreating thrombus *in silico* through bottom-up approaches (Fogelson & Neeves 2015, Kim et al 2019). Most *in silico* models so far neglect VWF (Fogelson & Neeves 2015, Pivkin et al 2006, Yazdani et al 2017) or incorporate the effect of VWF empirically (Mehrabadi et al 2016a). Multiscale computational methods (Chen et al 2013, Liu et al 2019c) that directly resolve the dynamics and interactions of molecules and cells hold the promise to model SIPA from first principles. To produce clinically relevant results, these models are yet to incorporate the real GP1b-A1 kinetic rates that are recently measured through single-molecule approaches (Chen et al 2019, Fu et al 2017, Kim et al 2010, Zhang et al 2015).

In this work, we develop a multiscale computational model informed by recent single-molecule measurements to recreate SIPA *in silico*. A new stochastic binding model for platelet-VWF GP1b-A1 bond formation is developed by combining the single-molecule kinetics measurement with the classical kinetic theory. The suspension dynamics of each individual platelet and VWF multimer are resolved directly through a multiscale blood flow solver (Liu et al 2019b, Liu et al 2019c, Liu et al 2018, Liu et al 2020, Mehrabadi et al 2016b, Reasor et al 2013). This *in silico* tool allows for studying SIPA with explicit molecular and cellular information directly informed by experimental measurements and may lead to new mechanisms of SIPA that are inaccessible to single-molecule measurements or micro-/macro-fluidic assays alone.

## 2. Methods

The computational method addresses the growth of an arterial thrombus under elevated high shear ($>10,000\ s^{-1}$), as shown in Fig. 1(a). The three necessary ingredients for arterial thrombosis (Casa et al 2015, Kim et al 2019) are included in the model system, as shown in Fig. 1(b). The red blood cells (RBCs) are neglected, considering platelet margination (Liu et al 2019b, Reasor et al 2013, Zhao & Shaqfeh 2011) and their secondary role in SIPA (Casa et al 2016). The thrombus-blood interface is simplified as a three-dimensional shear layer, where the bottom surface is assumed to be pre-adhered with VWFs as a pro-thrombotic surface. The schematic of the entire system is shown in Fig. 1(c), where the spatial and temporal dynamics of platelets and VWF polymers, the GP1b-A1 binding between VWF and platelet, the inter-/intra-association of VWF polymer, and the fluid-structure interactions are fully resolved. The methods to fully resolve the dynamics of platelets and VWF polymers is based on a multiscale particulate suspension flow method that is extensively validated against experiments (Liu et al 2019a, Liu et al 2019c, Liu et al 2018, Liu et al 2020), as described in the Supplementary Materials. Below we present a new platelet-VWF GP1b-A1 binding model informed by single-molecule measurements.

### 2.1. Kinetics-informed GP1b-A1 binding model

We have developed a binding model suitable for direct numerical simulation of ligand-receptor binding activity under hydrodynamic influence. Our theoretical approach combines the classic equilibrium binding kinetic theory (Bell 1978, Hammer & Apte 1992) with the recent single-molecule level kinetics measurement (Fu et al 2017, Kim et al 2010, Zhang et al 2015), enabling us to obtain the intrinsic GP1b-A1 kinetic rates that is decoupled from the transport effect. With the transport accounted directly through the suspension dynamics (see Supplementary Materials) and intrinsic kinetics given by the derived binding model, such computational system is generic in modeling ligand-receptor bindings subjected to hydrodynamic disturbances.

#### 2.1.1. GP1b-A1 on-rate

By quantifying the temporal binding percentage of the GP1b molecules onto a tethered VWF (Fu et al 2017), the apparent GP1b-A1 on-rate, $K_{on}$, was found to be VWF- tension dependent with a functional form,

$$K_{on} = \frac{K_{on}^m}{1+\exp\left[\frac{\Delta G - F_t \Delta x}{k_B T}\right]}, \quad (1)$$

where $F_t$ is the VWF internal tension force, $K_{on}^m$ is the maximum on-rate under sufficiently high VWF tension ($F_t \sim 100\ pN$), $\Delta G$ is the energy barrier between the no-tension state and the maximum-tension state, and $\Delta x$ is the displacement between the two states along the tension axis. The rate constants are given by Fu et al. as $K_{on}^m = 50.8 \pm 4.2 \times 10^6\ M^{-1}s^{-1}$, $\Delta G = 6.2 k_B T$, and $\Delta x = 1$ nm by fitting their experimental measurements.

It is noted that the apparent kinetic on-rate is controlled by both transport (convection and diffusion) and intrinsic association rate (Bell 1978), as shown in Fig, 2. Therefore, the apparent on-rate is not generic and depends on specific experimental settings. To obtain the intrinsic on-rate, the encounter and escape rates of the reacting molecules through convection and/or diffusion can be theoretically estimated and decoupled from the apparent on-rate.

Since the encounter and escape rates of GP1b and A1 molecules based on the experimental system by Fu et al. is diffusion-limited, the transport-induced GP1b-A1 encounter rate, $e_+$, and escape rate, $e_-$, can be estimated using statistical mechanics (Eigen 1974) as

$$e_+ = 4\pi(D_{GP1b} + D_{A1})R_{AB}, \tag{2}$$

$$e_- = 3(D_{GP1b} + D_{A1})R_{AB}^{-2}, \tag{3}$$

where $D_{GP1b}$ and $D_{A1}$ are the Brownian diffusion rate of GP1b molecules and A1 molecules, respectively. $R_{AB}$ is the encounter distance, which can be estimated as twice the radius of GP1b molecule, $R_{AB} \approx 2R_{GP1b}$. In the system of Fu et al., the VWF string (contains A1) was tethered whereas the GP1b molecules were suspended. Thus, the two diffusivities should satisfy $D_{GP1b} \gg D_{A1}$. The diffusion rate of GP1b can be quantified by its thermal diffusivity, i.e., $D_{GP1b} = k_B T/6\pi\mu R_{GP1b}$. The encounter and escape rates between GP1b and A1 can be approximated as

$$e_+ \approx 8\pi D_{GP1b} R_{GP1b} = \frac{4k_B T}{3\mu}, \tag{4}$$

$$e_- \approx 3D_{GP1b}/4R_{GP1b}^2 = \frac{k_B T}{8\pi\mu R_{GP1b}^3}, \tag{5}$$

Quantitatively, the above expressions yield an encounter rate of $e_+ = 3.5 \times 10^9 M^{-1} s^{-1}$ and an escape rate of $e_- = 7 \times 10^6 s^{-1}$.

Based on the measured apparent on-rate and the encounter and escape rates estimated above, the GPIb-A1 intrinsic on-rate, $k_+$, can be calculated according to (Bell 1978) as

$$k_+ = \frac{K_{on} e_-}{e_+ - K_{on}} \approx \frac{k_+^m}{1 + \exp\left[\frac{\Delta G - F_t \Delta x}{k_B T}\right]}, \tag{6}$$

where the maximum intrinsic on-rate is $k_+^m = K_{on}^m e_-/e_+ \approx 10^5 \ s^{-1}$. The $k_+$ value depends on the VWF tension force, $F_t$, which is explicitly tracked by the LB-LD method (Liu et al 2019c).

*2.1.2. GP1b-A1 off-rate*

The GP1b-A1 intrinsic off-rate quantifies the rupture dynamics of the GP1b-A1 bond, which is equivalent to the reciprocal of bond lifetime. The general functional form for the intrinsic GP1b-A1 off-rate follows the Bell-type equation (Bell 1978) as,

$$k_- = k_-^0 \exp\left[\frac{F_b \sigma}{k_B T}\right], \tag{7}$$

where $k_-^0$ is the zero-force intrinsic off-rate, and $\sigma$ is the reactive compliance that determines the degradation rate of the bond stability subjected to the GP1b-A1 bond force, $F_b$. The intrinsic GP1b-A1 off-rate, $k_-$, has been directly measured using single-molecule force-measurement assays (Chen et al 2019, Kim et al 2010, Zhang et al 2015). The rate constants used in the current studies are set to $k_-^0 \approx 1.0 \ s^{-1}$ and $\sigma = 0.2$ *nm* within the range of experimental measurements.

*2.1.3. Calculation of the bound/rupture probability*

With the obtained GPIb-A1 intrinsic rates, $k_+$ and $k_-$, the probabilities for GP1b-A1 bond formation and rupture can be determined. Through solving the master equation, $\frac{dP_i}{dt} = k_i(1 - P_i)$; $i \in \{+, -\}$ (Hammer & Apte 1992), the bound probability for two reactants (GP1b and A1) upon encounter (through transport), $P_+$, and the unbound probability of the existing GP1b-A1 bond, $P_-$, can be calculated according to,

$$P_+ = 1 - \exp[-k_+\Delta t], \quad (8)$$

$$P_- = 1 - \exp[-k_-\Delta t], \quad (9)$$

respectively, where $\Delta t$ is the time interval used to update the binding state. Because of the high temporal resolution in our computational method, $\Delta t$ is set to 100 LB timesteps, meaning the bound or rupture attempts between each VWF-A1 and platelet-GP1b are performed every 100 timesteps. The $P_+$ and $P_-$ are used to quantitatively regulate the likelihood of forming a new GP1b-A1 bond or rupturing an existing GP1b-A1 bond, respectively.

*2.2. Simulation setup and metrics*

The shear rate for the study is set to ~$O(10,000)$ $s^{-1}$, which represent typical shear levels for occlusive arterial thrombosis (Casa et al 2015) and match the shear rates studied in Ruggeri et al. (Ruggeri et al 2006). Platelet and soluble VWF (sVWF) at normal concentration have a volume fraction of $\phi_{plt} = 3.0\%$ and $\phi_{VWF} = 0.5\%$ (Chen et al 2013). In the cases comparing with (Ruggeri et al 2006), sVWF concentration is 2×normal and shear rate is 20,000 $s^{-1}$. At the thrombotic surface, immobilized VWF are tethered at one end in spots with a prescribed, physiologic surface density $S_{VWF}=0.1$ $ng/mm^2$ consistent with the experimental measurements (Hansen et al 2011, Moroi et al 1997). The contour length of the VWF multimer varies within the normal range 1.6~6.4 $\mu m$ (Fowler et al 1985). The computational domain has a dimension of 20×20×10 $\mu m^3$ along streamwise, shear-gradient and vorticity direction, respectively. Periodic boundary conditions are applied at the streamwise extrema of the computational domain. Simulations are initiated by seeding platelets and VWFs uniformly in the lower half of the domain to reduce the confinement effect of the upper moving boundary. VWF is initially set to globular equilibrium state. The shear rate is imposed from time zero. The motion and interactions of every single platelet and VWF are tracked seamlessly over time.

The SIPA process is quantified in terms of platelet agglomeration and capture. Agglomeration is the ability of the platelets to come together and form a larger solid mass that may travel in the flow or stick to the wall. Agglomeration is described by the platelet agglomeration index (PAI) as

$$\text{PAI} = \frac{2N_{cnt}}{N_{plt}}, \quad (10)$$

where $N_{cnt}$ denotes the number of pairs of platelets that are in close contact and $N_{plt}$ is the total number of platelets existed in the system. The criterion for platelet contact is defined such that the platelet-platelet separation is smaller than 10 nm.

Capture occurs when the translational velocity of platelet agglomerates become approximately zero, which is quantified through the platelet mobility index (PMI),

$$\text{PMI} = \frac{\langle V_{plt} \rangle}{\langle V_{fld} \rangle}, \quad (11)$$

where $\langle V_{plt} \rangle$ is the ensemble average of the platelet velocity in the system and $\langle V_{fld} \rangle$ denotes the volume-averaged fluid velocity.

## 3. Results

### 3.1. Dynamics of shear-induced platelet aggregation

Fig. 3 shows the time course of the SIPA (also see supplementary animation). VWF rapidly elongates in less than 0.5 ms. At t = 3 ms, platelet aggregates appear in the flow without being firmly captured. At t=7.5 ms, the platelet aggregates are captured and immobilized onto the thrombotic surface.

Fig. 4 plots the SIPA process in terms of agglomeration and capture quantified by PAI and PMI. PAI rises quickly between 0-6 ms, and then plateaus when all the platelets in the domain are connected by sVWF. In contrast, PMI remains relatively constant between 0-4 ms, decreases after 4 ms, and then falls to zero at 8 ms as the agglomerate is firmly captured as an aggregate stuck on the wall. We denote the time for PMI plateaus near 0 as the capture time for the platelet aggregate. The distinct time courses for agglomeration and mobility processes suggests that platelets agglomerate in the flow before being captured at the wall.

### 3.2. Aggregate morphology upon capture

The agglomerates formed through the entanglement of VWF in the flow (Goto et al 1998) can be firmly captured onto the iVWF surface and form aggregates under elevated high shear (Ruggeri et al 2006), facilitating the rapid platelet accumulation and occlusion of the arterial stenosis under hyper-shear conditions (Griffin et al 2018, Li et al 2012, Para et al 2011). Fig 5(a) depicts the morphology of the mural aggregates obtained both *in vitro* (Ruggeri et al 2006) and *in silico* with normal plasma VWF under a shear rate of 20,000 $s^{-1}$. The platelet aggregate obtained from our *in silico* model shows qualitatively similar morphology and aspect ratio compared to that observed in the microfluidic experiments (Ruggeri et al 2006). In this particular case, both *in silico* and in vitro studies show an elongated morphology of aggregate with an aspect ratio around 4.

### 3.3. Agglomerate capture time

Several timescales relevant to SIPA under elevated high shear are compared in Fig. 5(b). Considering a typical stenotic section that is $L_{st}$ =1000 $\mu m$ long in the flow direction (McCarty et al 2016), the platelet transit time can be estimated as

$$t_{trans} = \frac{L_{st}}{\dot{\gamma} R_{plt}}, \qquad (12)$$

where the radius of platelet is $R_{plt}$=1 $\mu m$. The platelet transit time can be estimated as $t_{trans}$ =100 to 1 ms at shear rate $10^4$ to $10^6$ $s^{-1}$. The minimal shear-induced platelet activation time, $t_{act}$, reported in (Hellums 1994) is is much greater than the platelet transit time, $t_{act} \gg t_{trans}$ when shear rate is $\ll 10^6$ $s^{-1}$, as plotted in Fig. 5(b). Platelet activation may be possible within the millisecond platelet transit time when shear rate exceeds $10^6$ $s^{-1}$, while such hyper shear would likely lead to the rupture of thrombus with embolization. As derived in the current study, the minimal time for the formation of GP1b-A1 is $t_+^{min} \approx 1/k_+^m = 10\,\mu s$, which acts as a lower bound for the time required for the capture of platelets.

The temporal analysis above confirms SIPA under elevated shear could occur independent of the shear-induced platelet activation (SIPAct) (Ruggeri et al 2006), as the platelet transit time through the arterial stenosis is much shorter than the minimal time for SIPAct. Moreover, to capture platelet agglomerates at the stenotic section, the capture time needs to be shorter than platelet transit time. Quantitatively, the timescales for the formation and capture of platelet agglomerates through SIPA under the shear rates of $10^4$ to $10^6$ $s^{-1}$ ranges from $t_{cap}$= 0.01 to 10 ms, as denoted in the "capture time w/o activation" area in Fig. 5(b). Outside this area, no agglomerates can be captured either due to the exposure time being shorter than the minimal GP1b-A1 bond formation time ($\ll 10$ $\mu s$, marked as "No capture" in Fig. 5b) or due to the time required for capture being longer than the platelet transit time ($\gg 10$ $ms$, marked in light blue or yellow as "No capture w/o activation" or "No capture w activation", respectively, in Fig. 5b).

The capture time obtained based on our *in silico* approach is $t_{cap} \approx 7.5 \pm 1.5$ msec for the shear rate of 10,000 $s^{-1}$ and $5.0 \pm 3.0$ msec for 20,000 $s^{-1}$, where the standard deviation is introduced by varying VWF length from 1.6 to 6.4 $\mu m$. We mark these capture times on the Hellums-type plot in Fig. 5(b) (orange symbols), which fall in the "Capture time w/o activation" area, consistent with the temporal constraints defined by experimental observations.

*3.4. Agglomerate traveling distance*

As a result of the capture time, that platelet aggregates would not form right away at the entrance of a high shear zone until the agglomerates are captured after certain traveling distance. By integrating the average velocity of platelets over time, the agglomerate traveling distance can be calculated as

$$L_t = \int_0^{t_{cap}} \langle V_{plt} \rangle dt. \tag{13}$$

For the cases to compare against Ruggeri's experimental observation at 20,000 $s^{-1}$ shear rate, the agglomerate traveling distance is 240±70 *μm* from the simulation. This streamwise agglomerate traveling distance compares well with the observation of no aggregates at the beginning of the high shear zone for 290±110 *μm* observed *in vitro* (Ruggeri et al 2006), as demonstrated in Fig. 5(c). The good quantitative agreement between the *in vitro* and *in silico* observation in terms of the agglomerate traveling distance further adds validity to the developed SIPA model pertinent to elevated high-shear arterial thrombosis.

**4. Discussion**

The theoretically derived intrinsic on-rate for GP1b-A1 bond formation informed by the single-molecule measurements in the current work quantitatively describes the minimal time required for the GP1b-A1 bond formation being around $1/k_+^m \sim 10$ $\mu$s. Note that the GP1b-A1 bond formation time has not been directly measured so far due to its ultra-short timescale. However, it is consistent with the lower bound of the theoretical estimation in (Wellings & Ku 2012), where they showed the firm capture of a single platelet through sufficient GP1b-A1 bonds under shear rates of 100,000 to 10,000 $s^{-1}$ requires a binding time of 15 to 150 $\mu$s. It is also found that the minimal GP1b-A1 formation time is typically much shorter (~10 $\mu s$) than the bond lifetime (0.01~1 s) (Chen et al 2013, Zhang et al 2015) under excessive shear stresses. This may explain the rapid accumulation of GP1b-A1 bonds needed for agglomerating (Goto et al 1998) and capturing platelets under elevated high shear (>10,000 $s^{-1}$) (Wellings & Ku 2012).

Through the *in silico* model, we predict the micro-aggregate formation through VWF entanglement under elevated shear (~10,000 $s^{-1}$) occurs in the millisecond timescale (~$10^{-3}$ s). This is much shorter than the minimal time for shear-induced platelet activation (1−$10^4$ s) (Hellums 1994) and the typical GP1b-A1 bond lifetime (0.1−10 s) (Chen et al 2019, Kim et al 2010, Zhang et al 2015). Such rapid process is made possible through both the ultra-fast intrinsic on rate of GP1b-A1 (~$10^5 s^{-1}$) and the convective effect from the elevated high shear (>10,000 $s^{-1}$), as an effective synergy of both kinetics and transport that drives SIPA in milliseconds. Moreover, the millisecond-timescale SIPA satisfies the short platelet transit time through high-shear stenosis (also in millisecond timescale), which aids in the explanation of rapid platelet accumulation and the subsequent occlusion observed in single-pass arterial thrombosis in vitro models (Casa & Ku 2017, Ku & Flannery 2007, Para et al 2011).

The SIPA model developed here has limitations. The model does not consider the presence of ADAMTS13. Since ADAMTS13 requires minutes to reduce VWF size significantly (Kretz et al 2015), it is expected that the addition of ADAMTS13 have little effect on the millisecond-timescale SIPA modelled in this study. Owing to the mesoscopic nature of the computational method, the SIPA model is restricted to simulating the micro-aggregate formation observed in microfluidic thrombosis experiments (Griffin et al 2018, Li et al 2012). To transition to macroscale thrombosis that typically occurs in millimeter spatial scales and minutes in temporal scales (Wootton & Ku 1999), other multiscale strategy is needed. Nonetheless, the multiscale model developed herein describes the detailed dynamics of SIPA *in silico*, covering biophysical events and features for micro-thrombus formation over a wide range of length ($10^{-9}$ to $10^{-5}$ meters) and time scales ($10^{-9}$ to $10^{-3}$ seconds) that are too difficult to precisely resolve using existing *in vitro* or *in vivo* thrombosis assays.

## 5. Conclusion

In this work, we develop a multiscale computational model that incorporates the minimal, necessary ingredients at both molecular and cellular scales required for SIPA under elevated shear pertinent to occlusive arterial thrombosis.

By theoretically decomposing the macroscopic kinetics into the transport component and the intrinsic-kinetics component, a new platelet-VWF GP1b-A1 binding kinetic model is developed to specify the intrinsic rates of GP1b-A1 bond formation. The derived intrinsic rate specifies the minimal time requires to form a GP1b-A1 bond (~10 $\mu$s), which has not been experimentally measured given its ultra-short timescale. The developed SIPA model shows both qualitative and quantitative agreement with the experiment in terms of the spatiotemporal characteristics of the mural micro-aggregates. SIPA under elevated high shear (~10,000 1/s) is predicted to be as fast as 1-10 milliseconds comparable to the platelet transit time over the throat of pathologic arterial stenosis, and much shorter compared to the time for shear-induced platelet activation (1-$10^4$ s) (Hellums 1994).

The multiscale model for SIPA at elevated high shear provides a predictive *in silico* tool for studying micro-thrombus formation with explicitly specified VWF-platelet interactions via GP1b-A1. This is typically difficult to obtain through single-molecule and micro-/macro-fluidic assays only. New insights obtained from the model may lead to the discovery of novel VWF-targeting anti-thrombotic therapies.

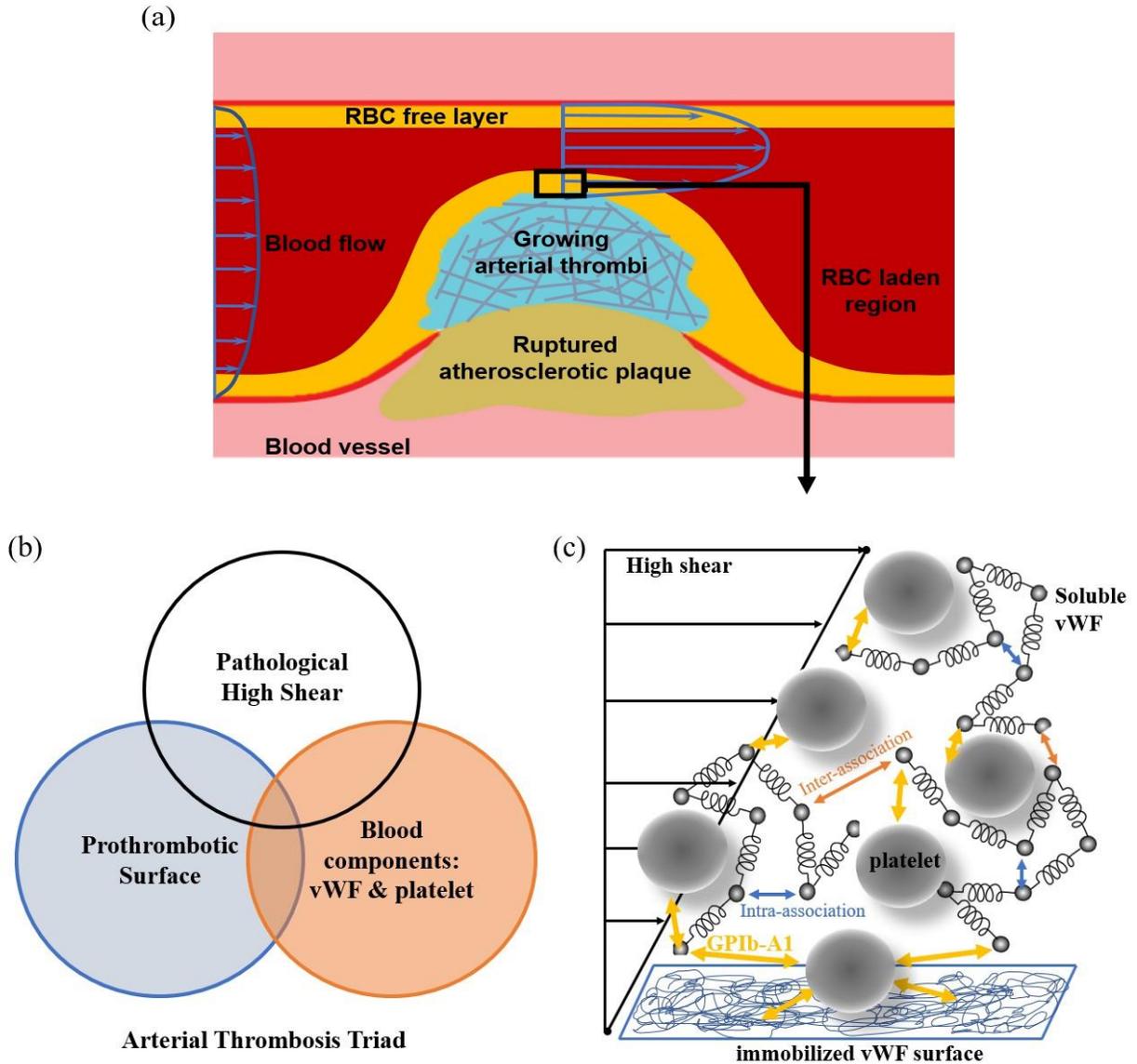

Figure 1: Schematics for the model system. (a) Blood flow through an arterial stenosis where an occlusive thrombus is forming and locally creating an elevated-high-shear environment. Further growth requires platelet aggregation under ultra-high shear stress. (b) The triad illustrates the necessary conditions for arterial thrombosis (Kim et al 2019). (c) The *in silico* model system represents a zoom-in view of the thrombi-blood interface subjected to elevated high shear. Shear-induced platelet aggregation (SIPA) occurs onto a VWF-rich prothrombotic surface. The SIPA process is supported by forming platelet-VWF GP1b-A1 bonds and the VWF inter-/intra-association bonds. Note (c) is a schematic representation of the least-required ingredients for SIPA and their bonding as shown in the arterial thrombosis triad in (b), not a snapshot from the simulation.

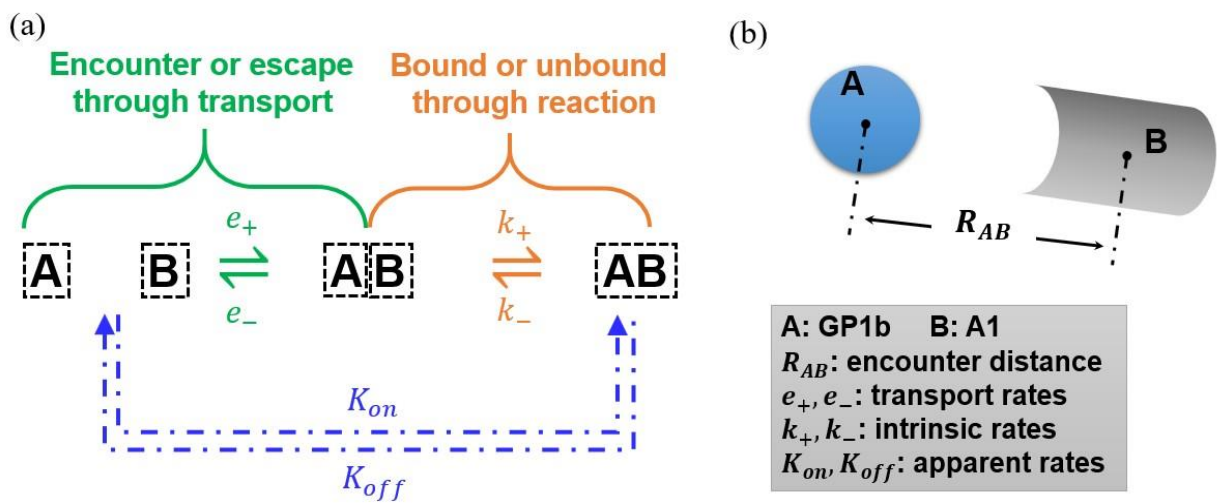

Figure 2: Schematic illustration of the ligand-receptor binding kinetics. (a) The apparent kinetic process can be decomposed into the transport portion and the intrinsic kinetics portion. The macroscopic association (dissociation) involves first the encounter (unbound) of the ligand and receptor, and then the bound (escape) of the two molecules. (b) The two reactive molecules (e.g., GP1b and A1) need to locate adjacent to each other within certain encounter distance, $R_{AB}$, in order to proceed with the subsequent intrinsic kinetic process.

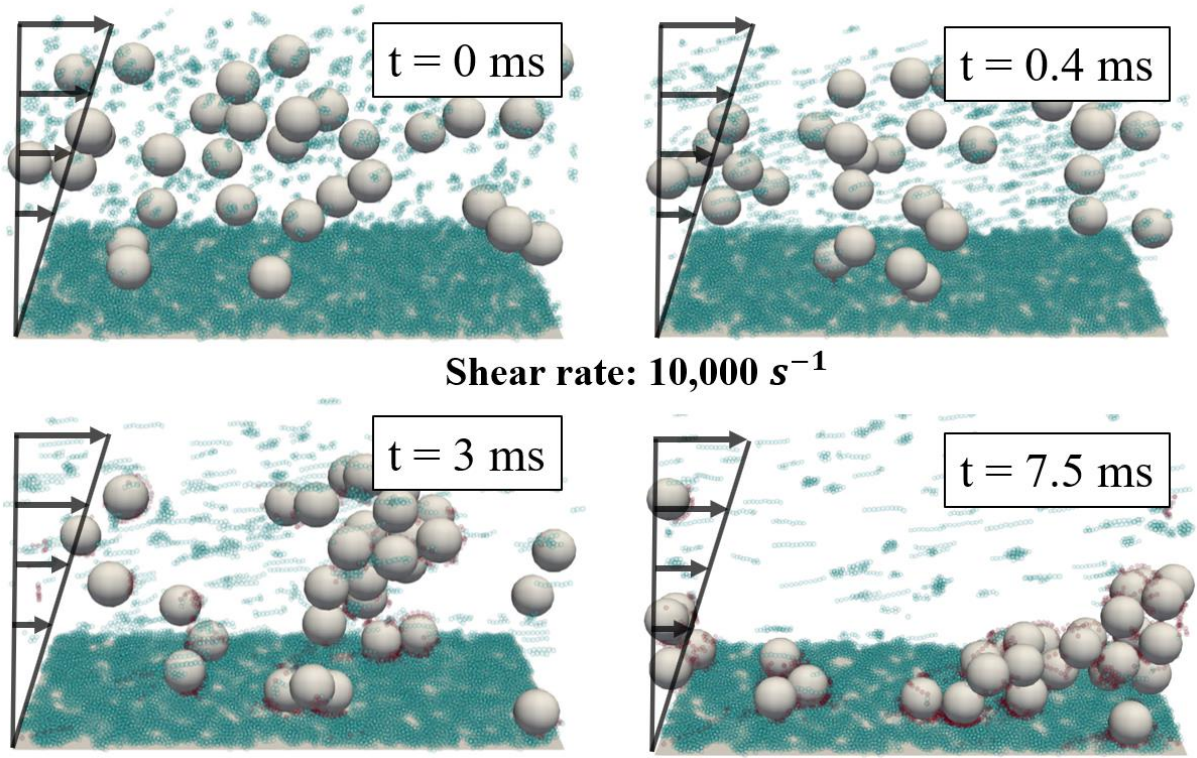

Figure 3: The dynamic process of SIPA specific to occlusive arterial thrombosis. The time sequence of the SIPA process under shear rate of 10,000 $s^{-1}$. The green beads are regular VWF dimers and red beads are activated VWF dimers (by forming GP1b-A1 bonds). In this particular case, platelet aggregates are formed in 8 ms, firmly adhered to the iVWF-covered thrombotic surface. The VWF has a contour length of 1.6 $\mu m$.

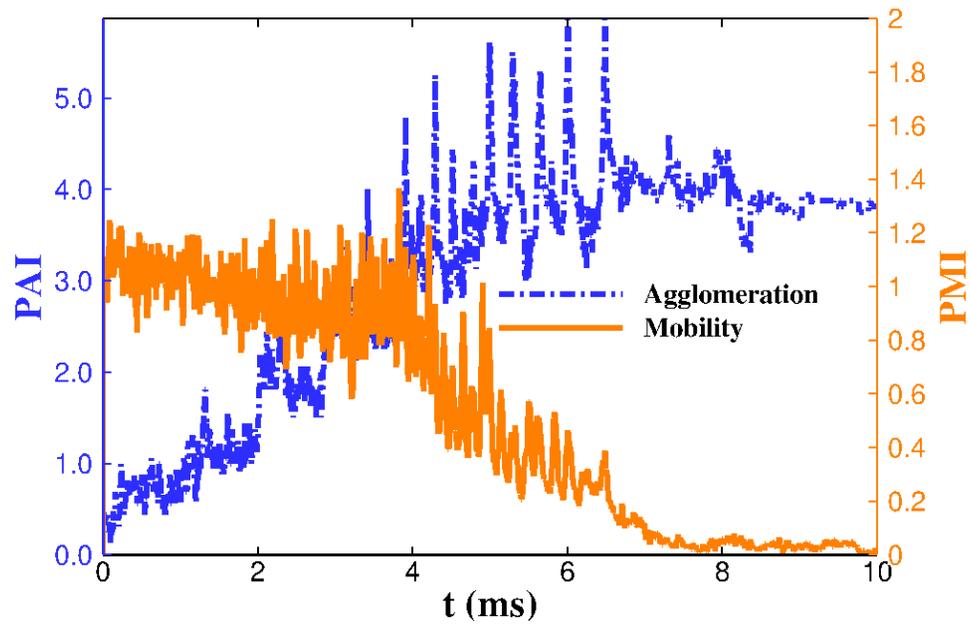

Figure 4: The platelet agglomeration index (PAI) and platelet mobility index (PMI) plotted against time. SIPA under elevated shear involves platelet agglomeration and capture processes.

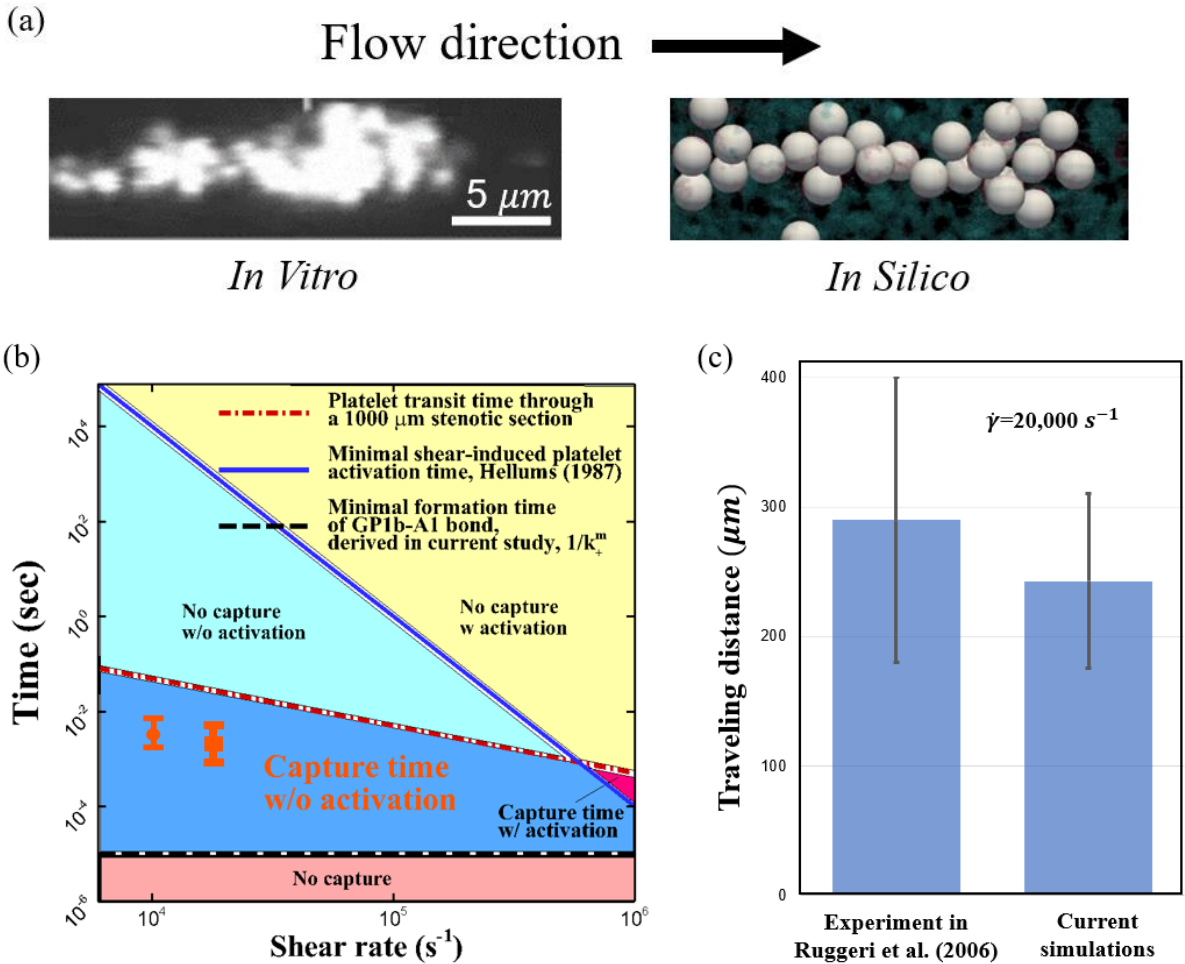

Figure 5: The morphological, temporal, and spatial characterization of SIPA. (a) The top-view morphology of the captured platelet agglomerates compared between the previous *in vitro* (left, adopted from Ruggeri et al. 2006) and the current *in silico* (right) results at the same scale. The capture agglomerates appear elongated due to the hydrodynamic drag forces. (b) Relevant time scales for SIPA. SIPA under elevated shear needs to be activation-independent and rapid, given the relatively short platelet transit time and long platelet activation time. The minimal shear-induced platelet activation time is adopted from Hellums et al. (1987) . Our derived maximum intrinsic on-rate for GP1b-A1 bond formation defines the minimal time required ($1/k_+^m$) for the capture of platelets. The simulation predicts the platelet aggregates formed at the wall in a few milliseconds (orange symbols) within the proposed blue area (capture time w/o activation) under shear rates of 10,000 $s^{-1}$ (circle) and 20,000 $s^{-1}$ (square). (c) The agglomerate traveling distance for SIPA under a shear rate of 20,000 $s^{-1}$ calculated based on the *in silico* model compares well with the *in vitro* measurements from Ruggeri et al. (2006).